\documentclass[draft]{amsart}
\usepackage{graphicx}
\usepackage{color}

\usepackage{amsmath,amssymb}
\usepackage{latexsym}

\newtheorem{theorem}{Theorem}[section]
\newtheorem{proposition}{Proposition}[section]

\theoremstyle{definition}

\theoremstyle{remark}

\numberwithin{equation}{section}

\newcommand{\R}{\mathbb R}

\newcommand{\Z}{\mathbb Z}

\newcommand{\imm}{\mathrm{i}}              
\newcommand{\e}{\mathrm{e}}                
\newcommand{\de}{\mathop{}\!\mathrm{d}}  
\newcommand{\pa}{\mathop{}\!\partial}

\newcommand{\ve}[1]{{\mathbf {#1}}}

\newcommand{\Cal}[1]{{\mathcal {#1}}}
\newcommand{\lr}[1]{\langle {#1} \rangle}
\newcommand{\tripla}{|\!|\!|}

\begin{document}

\title{Exponential dephasing of oscillators in the Kinetic Kuramoto Model}


\author{D. Benedetto}
\email{benedetto@mat.uniroma1.it}

\author{E. Caglioti}
\email{caglioti@mat.uniroma1.it}

\author{U. Montemagno}
\email{montemagno@mat.uniroma1.it}
\address{Dipartimento di Matematica, {\it Sapienza} Universit\`a di Roma - 
P.zle A. Moro, 5, 00185 Roma, Italy}

\subjclass[1991]{35A10, 35Q92, 74A25, 76N10, 92B25}

\keywords{Kuramoto model, dephasing, Landau damping, 
abstract Cauchy-Kowalewskaya theorem}

\date{\today}


\begin{abstract}
We study the kinetic Kuramoto model for coupled oscillators with
coupling constant below the synchronization threshold. 
We manage to prove that,
for any analytic initial datum, if the
interaction is small enough, the order parameter of the model
vanishes exponentially fast, and 
the solution is asymptotically described by a free flow.
This behavior is similar to the phenomenon of Landau damping in
plasma physics. 
In the proof we use a combination of techniques from Landau damping 
and from abstract Cauchy-Kowalewskaya theorem. 
\end{abstract}

\maketitle
\noindent

\section{Introduction}

The Kuramoto model is a mean-field model of coupled oscillators
proposed by Kuramoto to describe synchronization phenomena (see \cite{kuramoto},\cite{strogatz2000},\cite{bonilla}).
Any oscillator has a phase 
$\vartheta$, that can be considered defined mod $2\pi$, i.e.
in the one-dimensional torus 
$\Cal T$, and a ``natural frequency'' $\omega\in \R$.
In the kinetic limit the model reads as
\begin{equation}\label{noneq}
\begin{cases}
\partial_t f(t,\vartheta,\omega)+ \partial_\vartheta(v(t,\vartheta,\omega)f(t,\vartheta,\omega))=0\\
v(t,\vartheta,\omega)=\omega -\mu\int_{\mathcal{T}\times \R} \sin(\vartheta-\vartheta')f(t,\vartheta',\omega')\de \vartheta'\de \omega'\\
\end{cases},
\end{equation}
where $f(t,\vartheta,\omega)$ is a probability density in $\Cal T \times \R$,
$\int_{\mathcal{T}\times \R} 
\sin(\vartheta-\vartheta')f(t,\vartheta',\omega')\de \vartheta'\de \omega$
is the mean field interaction term, and $\mu>0$ is the coupling constant.
The distribution of the natural frequencies 
is $g(\omega ) = \int_{\Cal T}  f(t,\vartheta, \omega)\de\vartheta$,
which is a conserved quantity.

It can be useful to
represent the system \eqref{noneq} in the unitary circle
of the complex plane
by considering the oscillators as particles with position $\e^{\imm \vartheta}$.
The center of mass  is in the point
\begin{align}\label{orderparam}
&R(t)\e^{\imm \varphi(t)}=\int f(t,\vartheta,\omega)\e^{\imm \vartheta}\de \vartheta \de \omega.
\end{align}
$R$ and $\varphi$ are the ``order parameters'' of the model.
By this notation the coupling term can be rewritten as 
\begin{equation}
\label{eq:rriscritto}
\int_{\mathcal{T}\times \R} \sin(\vartheta-\vartheta')f(t,\vartheta',\omega')\de \vartheta'\de \omega' = R(t)\sin ( \vartheta - \varphi(t));
\end{equation}
so that the mean field interaction between the particles can be read as an attraction towards the phase of the center of mass, moduled by $R(t)$.

Existence and uniqueness results for the system \eqref{noneq} are obtained in
\cite{lancellotti}, where the \eqref{noneq} is rigorously derived by
doing the kinetic limit of the particle model introduce by Kuramoto.

The model has been intensively studied in the case where $g$ 
has compact support and the coupling
constant $\mu$ is sufficiently large to observe the asymptotic
synchronization
phenomenon (see \cite{MR2897541},\cite{carrillo},\cite{HHK}
and \cite{kuramoto2014},\cite{dong} for the
simpler case of $g(\omega)$ being a Dirac delta).
The asymptotic behavior of the solutions can be more 
complex: in particular, 
the total synchronization is impossible if $g$ has not compact support, 
and in this case it is expected a partial synchronization 
(see \cite{strogatz2000}). 

On the other hand, a completely different asymptotic behavior
is suggested in \cite{ldstrogatz},  in which, 
for small value of $\mu$,  it is shown a Landau-damping type results 
for a linearized model: the order parameter of a perturbation 
of the constant phase density relaxes to zero as the time 
goes to infinity, and the solution becomes an ``incoherent state''.
This behavior has been proved 
in the recent work \cite{giacomin} for the non linear model, 
where a full Landau-damping type result is obtained 
in the case of Sobolev regularity, showing that $R$ 
vanishes polynomially in time.

In this work we show, for small values of $\mu$, 
the asymptotic dephasing of the 
solutions of the nonlinear equation \eqref{noneq} in the case of
analytic initial data; in particular,
we prove that $R(t)$ vanishes exponentially
fast.
The techniques we use  take
inspiration from the works on Landau damping for the Vlasov equation
(see \cite{ldexp},\cite{villani}) 
and on the damping for the Euler equations
(see \cite{bedrossian}),  
and mostly from the more recent work \cite{ldsobolev}, in which 
it is shown the Landau damping in Sobolev norm for
the Vlasov HMF model.
Unlike the case of Sobolev regularity, the estimates 
in analytic norms do not close, so we adapt to this case
the abstract Cauchy-Kowalewskaya Theorem techniques, in a way
that allow us to obtain 
globally in time existence, while the  abstract Cauchy-Kowalewskaya
Theorem gives only finite time existence.

We remark that the phenomenon of dephasing of the solutions 
of the Kuramoto model has been studied 
in the recent work \cite{giacomin}, in which it is shown 
a Landau-damping type result for eq. \eqref{noneq} in Sobolev norms,
and in the work \cite{dephback} in which it is shown the existence 
of dephasing solutions with a prescribed asymptotic behavior, with  
analytical or Sobolev regularity.

\section{Dephasing}\label{sez:forward}

In terms of the order parameters, the 
kinetic Kuramoto equation reads as
\begin{equation}\label{noneq2}
\begin{cases}
\partial_t f(t,\vartheta,\omega)+ \partial_\vartheta(v(t,\vartheta,\omega)f(t,\vartheta,\omega))=0\\
v(t,\vartheta,\omega)=\omega -\mu R(t)\sin(\vartheta-\varphi(t))\\
R(t)\e^{\imm \varphi(t)}=\int_{\Cal T\times \R} f(t,\vartheta,\omega)\e^{\imm \vartheta}\de \vartheta \de \omega.
\end{cases}.
\end{equation}
If $\mu=0$, the solutions of eq. \eqref{noneq2} is 
$$f(t,\vartheta,\omega) = f_0(\vartheta - \omega t, \omega)$$
where $f_0$  is the initial datum.
Our aim is to show that, 
if $f_0(\vartheta,\omega)$ is bounded in some analytical norm,
and $\mu >0$ is sufficiently small, 
there exists an asymptotic state $h_\infty$ such that
$$f(t,\vartheta+\omega t,\omega) \to h_{\infty}(\vartheta,\omega)$$
exponentially fast.
In other words,
the solutions $f(t,\vartheta,\omega)$ 
asymtotically approches the  incoherent state
$h_{\infty}(\vartheta - \omega t,\omega)$, 
i.e. a function 
transported by the free flow 
In this sense, the solutions  show 
an asymptotic dephasing.
The key ingredient of the proof 
is the exponential decay of the order parameter
$R(t)$, which, as noted in \cite{mirolloor}, 
can only be obtained for analytic initial data.

To state precisely our result, we define 
$h(t,\vartheta,\omega) = f(t,\vartheta + \omega t, \omega)$,
with initial datum 
$h(0,\vartheta,\omega) = f_0(\vartheta,\omega)$ and which verifies,
from eq. \eqref{noneq2}, the equation
\begin{equation}\label{eqh}
\pa_t h(t,\vartheta,\omega) 
= - \mu  R(t) \pa_\vartheta \left( \sin(\vartheta  
+\omega t - \varphi(t)
h(t,\vartheta,\omega)\right).
\end{equation}
Defining for $k\in \Z$ and $\eta\in \R$ 
$$\hat{h}_k(t,\eta) = \frac{1}{{2\pi}} 
\int_{\Cal T\times \R} \de \vartheta \de \omega
h(t,\vartheta,\omega) \e^{-\imm 
\vartheta k -\imm \omega \eta}$$
eq. \eqref{eqh} in Fourier space is
\begin{align}\label{eqhfourier}
&\pa_t\hat{h}_k(t,\eta)= \mu {\widehat{\ve L_t h}} (k,\eta), \ \ \text{where}
\\
\label{eq:L}
&\widehat{\ve L_t f}(k,\eta) \doteq
k\sum_{m=\pm 1}\frac{m}{2}  z_m(t)\hat {g}_{k-m}(\eta-mt),
\end{align}
where the order parameters read as
$$z_{\pm 1}(t)=\hat{h}_{\pm 1}(t,\pm t)=R(t)\e^{\mp\imm \varphi(t)},\ 
|z(t)|=R(t).$$
Integrating in time eq.s \eqref{eqhfourier} we have
\begin{equation}
\label{eq:integrale}
\hat{h}_k(t,\eta) = \hat{h}_k(0,\eta)  + \mu k 
\sum_{m\in \pm 1} \frac{m}{2} \int_0^t z_m(s)\hat {h}_{k-m}(s,\eta-ms),
\de s.
\end{equation}
In the sequel, we consider separately the evolution in time 
of the order parameter $z_{\pm 1}$, and of the function $h$; in this
sense $\ve L_t$ can be considered a linear operator in $h$.

We use analytical norms for $h$ so, to make the notation lighter, we give the following definitions:
\begin{equation}
\langle t\rangle =( 1+t^2)^{\frac{1}{2}},
\quad \langle k,\eta \rangle =( 1+k^2+\eta^2)^{\frac{1}{2}};
\end{equation}
note that $\langle \cdot \rangle$ verifies the triangular inequality, 
as follows from easy calculation:
\begin{equation}
\langle k_1 + k_2 , \eta_1 +\eta_2 \rangle \le 
\langle k_1 , \eta_1 \rangle +
\langle k_2 , \eta_2 \rangle.
\end{equation}
For $\lambda,p\ge 0$, we define the weight
\begin{equation}\label{eq:A}
A^{\lambda,p}_k(\eta)=
\e^{\lambda \langle k,\eta\rangle}\langle k,\eta\rangle^{p},
\end{equation}
and the norms:
\begin{equation}
\label{norm}
\|f\|_{\lambda,p} =
\sup_{k\in \Z, \eta \in \R}
A^{\lambda,p}_k(\eta) |\hat f_k(\eta)|. 
\end{equation}
We call $\Cal X_{\lambda,p}$ the space of function $f$ with 
finite $\|f\|_{\lambda,p}$ norm.

Using this norm, it is easy to show that if $\mu=0$,
we obtain the exponential decay of the order parameter 
$z_{\pm 1}(t)$. Let us first
obtain an equation for $z_1=R(t)\e^{-\imm \varphi(t)}$ 
setting $k=1$ and $\eta=t$ in \eqref{eq:integrale} 
\begin{equation}\label{eq:z1}
z_1(t)=\hat{h}_1(0,t)+\mu \sum_{m=\pm 1}\frac{m}{2} \int_{0}^t z_m(s)
\hat {h}_{1-m}(s,t-ms)\de s,
\end{equation}
where $h(0,\vartheta,\eta)$ is the initial datum $f_0(\vartheta,\eta)$.
Choosing $\lambda,p\ge 0$, 
the first term, due to the  free flow, is bounded by
$$\|f_0\|_{\lambda,p} \e^{-\lambda \lr{0,t} }\lr{0,t}^p 
\le C\e^{-\lambda t}\lr{t}^p \|f_0\|_{\lambda,p}$$
where, here and in the following, $C$ is a suitable
time independent constant.
This estimate suggests that we can control the quantities
\begin{equation}
\label{eq:r}
r_{\lambda,p}(t) = |z_{\pm 1}(t)| \e^{\lambda t} \lr{t}^p.
\end{equation}
An uniform in time  bound of $r_{\lambda,p}(t)$ is equivalent to an 
exponential decay of $R(t)$. The aim of this paper is to show that, if $\mu$ is sufficiently small,
the other terms in \eqref{eq:z1} do not prevent a uniform 
estimate for $r_{\lambda,p}$.

\section{A-priori estimates}

In the case $\mu>0$, 
$r_{\lambda,p}(t)$ can be estimated 
as in the following proposition.
\begin{proposition}
\label{propo:stimar}
For $\lambda,p\ge 0$ 
\begin{equation}
\label{eq:stimar}
\begin{aligned}
r_{\lambda,p}(t) \le &C \|f_0\|_{\lambda,p} \\
&\!+ \mu C 
\|f_0\|_{\lambda,p} \int_0^t\!\! r_{\lambda,p}(s) 
\left( \frac 1{\lr{s}^p}+\frac 1{\lr{t-s}^p}
\right)ds + \mu C \int_0^t\!\! r_{\lambda,p}(s) 
\frac{ \|h(s)\|_{\lambda,p}}{\lr{s}^p}ds.
\end{aligned}
\end{equation}
\end{proposition}
\proof

The first term of the r.h.s. of \eqref{eq:z1}
is simply bounded by 
$$\|f_0\|_{\lambda,p} \e^{\lambda \lr{0,t} }\lr{0,t}^p 
\le C\e^{\lambda t}\lr{t}^p \|f_0\|_{\lambda,p}.$$
Since $\hat{f}_0(t,\eta) = \hat{f}_0(0,\eta)$, the term with 
$m=1$ is 
\begin{equation}\nonumber 
\int_{0}^t z_1(s) 
\hat {f}_{0}(s,t-s)\de s = 
\int_{0}^t z_1(s)
\hat {f}_{0}(0,t-s)\de s,\end{equation}
which is bounded by
\begin{align}&C \|f_0\|_{\lambda,p}\int_{0}^t 
\e^{-\lambda \langle 0,t-s \rangle}  \langle t-s \rangle^{-p}
 |z_1(s)| \de s
\le
\nonumber
\\
&\le C  \|f_0\|_{\lambda,p}\int_0^{t} r_{\lambda,p}(s)
\e^{-\lambda s }  \langle s \rangle^{-p}
\e^{-\lambda (t-s)}  \langle t-s \rangle^{-p}\de s,
\label{stiR3}
\end{align}
where we used that $\lr{0,t-s}\ge (t-s)$.
Multiplying by $\e^{\lambda t} \lr{t}^{p}$,
we have the estimate
$$\int_0^t  r_{\lambda,p}(s) \frac { \lr{t}^p}
{\langle t-s \rangle^{p}  \langle s \rangle^{p}}\de s
\le 
\int_0^t  r_{\lambda,p}(s) \left( 
\frac 1{\lr{s}^p} + 
\frac 1{\langle t-s \rangle^{p}}\right) \de s,$$
because $\lr{t}^{p} \le C(\lr{s}^{p} + \lr{t-s}^{p})$.

The term with $m=-1$ is bounded by
\begin{equation}\label{stiR5}
\int_0^t |z_{\pm 1}(s)| \,|h_2(s,t+s)| \de s \le 
\int_0^{t} r_{\lambda,p}(s) \|h(s)\|_{\lambda,p} 
\e^{-\lambda s - \lambda \lr {t+s}} {\lr{s}^{-p} \lr{t+s}^{-p}}
\de s.\end{equation}
We conclude the estimate \eqref{eq:stimar} by multiplying 
by $\e^{\lambda t} \lr{t}^{p}$ and noting
that $\lr{t}^{p} \le \lr{t+s}^{p}$.
\endproof

In order to estimate $\|h\|_{\lambda,p}$, 
we need to control its time derivative.
\begin{proposition}
 \label{prop:stimaL}
Given $z_{\pm 1}(t)$,
for $\lambda, p\ge 0$, 
$\ve L_t$ is a continuous  operator from 
from $\Cal X_{\lambda,p+1}$ to $\Cal X_{\lambda,p}$:
\begin{equation}
 \label{eq:stimap+1}
\|\ve L_t h(t)\|_{\lambda,p} \le C 
\left( r_{\lambda,0}(t) \|h(t)\|_{\lambda,p+1} + 
r_{\lambda,p}(t) \|h(t)\|_{\lambda,1}\right).
\end{equation}
$\ve L_t$ is also continuous 
from $\Cal X_{\lambda',p}$ to $\Cal X_{\lambda,p}$,
when $\lambda'>\lambda$, in fact
\begin{equation}
\label{eq:pl}
\|f\|_{\lambda,p+1} \le \frac {1}{\lambda'-\lambda}\|f\|_{\lambda',p}.
\end{equation}
\end{proposition}

\proof

Recalling the definition of $A^{\lambda,p} $ in 
\eqref{eq:A}, we write
$$A^{\lambda,p}_k(\eta) |\ve L_t h (k,\eta)| \le 
\frac 12 |z_1(t)|\, |k| \sum_{m=\pm 1}
\e^{\lambda \lr{k,\eta}} \lr{k,\eta}^p 
  |\hat {h}_{k-m}(t,\eta-mt)|;$$
then, by the triangular inequality,
$$\lr{k,\eta}\le \lr{k-m,\eta -mt}  + \lr{m,mt},$$
and that, when $|m|=1$,
$\lr{m,mt} \le C + t$,
we have
$$\e^{\lambda \lr{k,\eta}} \le C \e^{\lambda t} \e^{\lambda 
\lr {k-m,\eta - m t}}.$$
Since $\lr{m,mt}$ also verifies $\lr{m,mt} \le Ct$, it is true that
$$\lr{k,\eta}^p\le C\left( \lr{k-m,\eta-mt}^p + \lr{t}^p\right),$$
which implies
$$
\begin{aligned}
&A^{\lambda,p}_k(\eta) |\ve L_t h (k,\eta)| \le\\
&\le C\e^{\lambda t} R(t) 
\sum_{m=\pm 1} \e^{\lambda \lr{k-m,\eta - mt}} 
 |k| \lr{k-m,\eta-mt}^p  |\hat{h}_{k-m}(\eta - mt)| + \\
&\quad +C\e^{\lambda t} R(t) 
\sum_{m=\pm 1} \e^{\lambda \lr{k-m,\eta - mt}} |k| \lr{t}^p |\hat{h}_{k-m}(\eta - mt)|.
\end{aligned}$$
Using that $|k|\le \lr{k-m,\eta-mt}$, we obtain the thesis
estimating the  first term 
with
$$Cr_{\lambda,0}(t) \|h(t)\|_{\lambda,p+1}$$
and the second with
$$Cr_{\lambda,p}(t) \|h(t)\|_{\lambda,1}.$$
\endproof

Let us discuss how to choose the norms that will 
allow us to obtain closed estimates for $h$ and $z_{\pm 1}$.
In the Landau Damping type results in the case of 
Sobolev regularity of order $\gamma$, the choice of a 
suitable Hilbert space  ${\Cal H}_\gamma$, with norm $\|h\|_{\Cal H_\gamma}$, 
guarantees that $L_t$ is a continuous map from ${\Cal H}_\gamma$ 
in the same ${\Cal H}_\gamma$ (see \cite{giacomin}).
Then the results are achieved estimating, globally in time, 
a term of the type 
$\|h(t)\|_{\Cal H_\gamma}/\lr{t}$, for suitable values of $\gamma$, 
and, correspondingly, $\lr{t}^{\gamma}R(t)$.
In the case of analytical regularity we can not obtain this behavior
and we have to take into account that 
in \eqref{eq:stimap+1} we can only estimate  $\ve L_t h$ 
in a norm that is weaker than the one of $h$.
We give closed estimates by mixing 
the typical norms used in Landau-Damping type results 
with the norms needed for the
proof of  the abstract 
Cauchy-Kowalewskaya theorem, following
in particular  \cite{caflish}.

Given $\lambda_0 > 0 $ and $a$ such that $0<a<2\lambda_0/\pi$, 
for $t\ge 0$ and $\lambda < \lambda_0$, we define the weight
\begin{equation}
\label{eq:beta} 
\beta(t,\lambda) = 
\beta_a(t,\lambda) = \lambda_0 - \lambda - a\int_0^t \frac {\de s}{\lr{s}^2} =
\lambda_0 - \lambda - a \arctan t.
\end{equation}
This function is positive for decreasing in time values of $\lambda$, 
and, as in abstract Cauchy-Kowalewskaya theorems, 
we use it to taking into account the loss of analytical 
regularity due to the spatial derivative in  
$\vartheta$ in the operator $\ve L_t$. 
In \cite{caflish} and 
in other proofs of the abstract Cauchy-Kowalewskaya Theorem,
the time dependence of the weight is linear and the solutions
exists only for finite time. Here the
 Landau-Damping type estimates allow us to choose a weight convergent 
in time, which, if  $a<2\lambda_0/\pi$ 
give the analyticity also for $t\to +\infty$.

More precisely, we define the Banach space 
$\tilde{\Cal B}_{a,p}$ as the space
of the functions $h(t)$ such that, if $\beta(t,\lambda) >0$, 
$h(t)\in \Cal X_{\lambda,p}$. The norm in $\tilde{\Cal B}_{a,p}$ is
\begin{equation}
\tripla h\tripla_{a,p} = \sup_{\lambda,t: \beta(t,\lambda)>0} 
\beta^{1/2}(t,\lambda)
\|h\|_{\lambda,p}. 
\end{equation}
Finally, fixing $\gamma\ge 3$, we define the norm
\begin{equation}
\label{eq:normaa}
\tripla h\tripla_a = 
\tripla h \tripla_{a,1} + \tripla h(\cdot)/\lr{\cdot}\tripla_{a,\gamma},
\end{equation}
and the corresponding Banach space $\Cal B_a$, of the function $h$
with $\tripla h \tripla_a$ bounded.
With little abuse of notation, we write:
\begin{equation}
\label{eq:normaaR}
\tripla R\tripla_a = \sup_{\lambda,t: \beta(t,\lambda)>0} r_{\lambda,\gamma}(t)
 = \sup_{\lambda,t: \beta(t,\lambda)>0} R(t) \e^{\lambda t} \lr{t}^{\gamma}
\end{equation}
Now, we prove the a-priori estimates in $\Cal B_{a}$ 
which allow us to construct the solutions. 
\begin{proposition}
\label{propo:stimah}
Given $z_{\pm 1}(t)$ 
with $\tripla R \tripla_a < +\infty$, 
if $h=h(t,\vartheta,\eta)$ solves eq. \eqref{eqhfourier}
then it satisfies
\begin{equation}
\tripla h \tripla_a \le C\|f_0\|_{\lambda_0,\gamma} + 
C{\mu} \tripla R\tripla_a \, \tripla h\tripla_a.
\end{equation}
\end{proposition}
\proof
First we estimate $\|h(t)\|_{\lambda,1}$, for $\lambda$ such that
$\beta(t,\lambda)>0$.
Using \eqref{eq:integrale} and the estimate \eqref{eq:stimap+1}
with $p=1$, 
we have
\begin{equation}
\label{eq:h1}
\|h(t)\|_{\lambda,1} \le C\|f_0\|_{\lambda_0} + 
C\mu \int_0^t r_{\lambda,\gamma}(s) \left( 
\frac 1{\lr{s}^{\gamma-1 }} \frac {\|h(s)\|_{\lambda,\gamma}}{\lr{s}} + 
\frac 1{\lr{s}^{\gamma-1}} \|h(s)\|_{\lambda,1}\right)\de s.
\end{equation}
Multiplying by $\beta^{1/2}(t,\lambda)$:
\begin{equation}
\label{eq:h1beta}
\beta^{1/2}(t,\lambda)\|h(t)\|_{\lambda,1} \le C
\|f_0\|_{\lambda_0} + 
C\mu  \tripla h\tripla_a \, \tripla R\tripla_a
\int_0^t  \frac 1{\lr{s}^{2}} \frac {\beta^{1/2}(t,\lambda)}
{\beta^{1/2}(s,\lambda)}\de s,
\end{equation}
where we have used that for $\gamma \ge 3$, $\lr{s}^{\gamma -1 } \ge 
\lr{s}^{2}$. Using that $\beta(t) \le \beta(s)$ we estimate 
the time integral with a constant, then
$$\tripla h \tripla_{a,1} \le C \|f_0\|_{\lambda_0,\gamma} + C\mu \tripla R
\tripla_a \, \tripla h \tripla_a.
$$
Now we estimate $\|h\|_{\lambda,\gamma}$: 
using eq. \eqref{eq:integrale} and the estimates
 \eqref{eq:stimap+1}, \eqref{eq:pl} with $p=\gamma$
\begin{equation}
\label{eq:hgamma}
\|h(t)\|_{\lambda,\gamma} \le C \|f_0\|_{\lambda_0,\gamma} + 
C\mu \int_0^t r_{\lambda,\gamma}(s) \left( \frac 1{\lr{s}^{\gamma}} 
\frac {\|h(s)\|_{\lambda'(s),\gamma}}{\lambda'(s)-\lambda} + 
\|h(s)\|_{\lambda,1} \right) \de s,
\end{equation}
for any $\lambda'(s)>\lambda$ 
such that $\lambda_0 - \lambda'(s) - a \arctan (s) > 0$.
Dividing by $\lr{t}$ and multiplying by $\beta^{1/2}(t,\lambda)$,
we obtain
\begin{equation}
\frac {\beta^{1/2}(t,\lambda)}{\lr{t}}\|h\|_{\lambda,\gamma}
\le  C\|f_0\|_{\lambda_0,\gamma} + 
C\mu  \tripla h\tripla_a \, \tripla R\tripla_a (I_1+I_2),
\end{equation}
where
\begin{align}
&I_1 = \frac {\beta^{1/2}(t,\lambda)}{\lr{t}}
\int_0^t \frac {\de s}{\lr{s}^{2}\beta^{1/2}(s,\lambda) (\lambda'(s)-\lambda)},
\nonumber\\
&I_2 = \frac {1}{\lr{t}}
\int_0^t \frac {\beta^{1/2}(t,\lambda)}{\beta^{1/2}(s,\lambda)} \de s.
\nonumber
\end{align}
$I_2$ is less than a constant because $\beta(t,\lambda) \le 
\beta(s,\lambda)$, for $s\le t$. \\
In $I_1$, we chose $\lambda'= \lambda'(s) > \lambda$ as
$$\lambda'(s) = \frac 12 (\lambda_0 - a \arctan s) + \frac {\lambda}2,$$
which verifies
$$\beta(s,\lambda'(s)) = \frac 12 \beta(s,\lambda)>0,$$
and
$$\lambda'(s) - \lambda = \frac 12 \beta(s)\ge \frac 12 \beta(t) > 0.$$
Then $I_1$ is bounded by 
$$I_1 \le 2\frac {\beta^{1/2}(t,\lambda)}{\lr{t}}
\int_0^t \frac {\de s}{\lr{s}^{2}\beta(s,\lambda)^{3/2}}.$$
Since $\de \beta/ \de s = -a /\lr{s}^2$, 
the time integral can be explicitly computed 
and gives:
$$
\int_0^t \frac {\de s}{\lr{s}^{2}\beta(s)^{3/2}} 
= \frac 2a  \left( \frac 1{\beta^{1/2}(t,\lambda)}
- \frac 1{\beta^{1/2}(0,\lambda)} \right),$$
then also $I_1$ is less then a constant.
\endproof

Now we estimate $\tripla R \tripla_a$.

\begin{proposition}
\label{propo:stimaz1}
Fixed $h$ such that $\tripla h\tripla_a < +\infty$,  
if $z_{\pm 1}(t)=R(t) \e^{\mp \imm \varphi(t)}$ solves \eqref{eq:z1}, then 
\begin{equation}
\label{eq:stimaR}
\tripla R \tripla_a \le C \|f_0\|_{\lambda_0, \gamma} 
( 1+ \mu \tripla R \tripla_a ) +  C \mu \tripla R \tripla_a \, 
\tripla h \tripla_a. 
\end{equation}
\end{proposition}
\proof 
We use \eqref{eq:stimar} with $p=\gamma$: the estimate of the first
two terms are obvious; the last one is bounded by 
$\mu C \tripla R \tripla_a \, \tripla h \tripla_a $ times the integral
$$\int_0^t \frac {\de s}{\lr{s}^2 \beta^{1/2}(s,\lambda)}= 
\frac 2a \left( \beta^{1/2}(0,\lambda)-\beta^{1/2}(t,\lambda)\right) \le 
\frac 2a \lambda_0^{1/2}$$
\endproof

\section{The main theorem}

\begin{theorem}
For $\lambda_0>0$ and $\gamma \ge 3$, 
if $\|f_0\|_{\lambda_0,\gamma}$ is bounded, 
for $\mu$ sufficiently small, the unique solution  $h(t,\vartheta,\omega)$ 
of \eqref{eqh} with initial datum $f_0(\vartheta,\omega)$ 
verifies $\tripla h \tripla_a < C$ and $\tripla R \tripla_a < C$. 

As a consequence, $R(t)\to 0$ exponentially fast and 
there exists $h_{\infty}(\vartheta, \omega)$ with 
$\|h_{\infty}\|_{\bar \lambda,\gamma} <+\infty$
for some $\bar \lambda>0$, 
such that
$$f(t,\vartheta+\omega t,\omega) = h(t,\vartheta,\omega)
\to h_{\infty}(\vartheta,\omega)$$
exponentially fast.
\end{theorem}
\proof

We construct the solution with an iterative procedure.
For $n\ge 0$
\begin{align}
\label{eq:h0}
&h^0(t,\vartheta,\omega) = f_0(\vartheta,\omega)\\ 
\label{eq:z1n}
&z_1^n(t)=\hat{h}^n_1(0,t)+\mu \sum_{m=\pm 1}\frac{m}{2} \int_{0}^t z^n_m(s)
\hat {h}^n_{1-m}(s,t-ms)\de s,\\
\label{eq:Ln}
&\widehat{\ve L_t^{n} f}(k,\eta) \doteq
k\sum_{m=\pm 1}\frac{m}{2}  z_m^{n}(t)\hat {f}_{k-m}(\eta-mt),\\
\label{eq:hn}
&\pa_t\hat{h}^{n+1}_k(t,\eta)= \mu \widehat{\ve L_t^{n} h^{n+1}}(k,\eta), 
\end{align}
where in eq. \eqref{eq:z1n} $z_{-1}^n$ is the conjugate of $z_1^n$.
The linear problems in eq. \eqref{eq:z1n} and in eq. \eqref{eq:hn}
are easily solvable, and the solutions verify the 
analogous of the a-priori estimate provided by 
Prop. \ref{propo:stimah} and Prop. \ref{propo:stimaz1}:
$$\tripla R^n \tripla_a  \le C \|f_0\|_{\lambda_0,\gamma}
\left( 1+ \mu \tripla R^n \tripla_a\right) + C \mu \tripla R^n \tripla_a\,
\tripla h^n\tripla_a$$ 
and 
$$\tripla h^{n+1} \tripla_a \le C \|f_0\|_{\lambda_0,\gamma}+ 
C \mu \tripla R^n\tripla_a\,\tripla h^{n+1} \tripla_a.$$
Using that $\tripla h^0\tripla_a \le C\|f_0\|_{\lambda_0,\gamma}$
we can inductively prove that, if $\mu \|f_0\|_{\lambda_0,\gamma}$
is sufficiently small, then
$$\tripla h^n\tripla_a \le C\|f_0\|_{\lambda_0,\gamma},\ \ \text{ and }
\tripla R^n\tripla_a \le C\|f_0\|_{\lambda_0,\gamma},
$$
uniformly in $n$. 
Choosing $a'>a$ with $a'<2\lambda_0/\pi$, 
and estimating the operator $\ve L_t^n$ as in Prop. \eqref{prop:stimaL}, 
we have that all the first derivative of $h^n$ are 
uniformly bounded in the region defined by $\beta_{a'}(t,\lambda) > 0$;
then, for subsequences, $h^n$ converges  to some 
$h\in \tilde{\Cal B}_{a',\gamma}$.
Correspondingly, $z^{n}_{\pm 1}$ converges to $z_{\pm 1}$ with 
$\tripla z_{\pm 1}\tripla_a$ bounded. 
The function $h$ and $z_{\pm 1} $ solve the coupled equations 
\eqref{eqhfourier} and \eqref{eq:z1}. Moreover, 
putting $k=1$ and $\eta=t$ in \eqref{eq:integrale} 
$$\hat h_1(t,t) 
= \hat h_1(0,t) + \mu \sum_{m= \pm 1}\int_0^t \frac m2 z_m(s) \hat h_{k-m}
(s,\eta - ms) \de s = z_1(t)$$
as follows form \eqref{eq:z1}.
Then $h$ solves the non linear equation \eqref{eqhfourier}, 
and its uniqueness is guarantees by the 
uniqueness of regular solutions (see \cite{lancellotti})
(note that the uniqueness implies the convergence to $h$ and $z$ 
for the full sequences $h_n$ and $z_n$).\\

Finally, 
let $\bar \lambda>0$, with $\bar {\lambda} < \lambda_0 - a'\pi/2$.
Then 
$$\|\ve L_t h\|_{0,\gamma} \le \frac C{\bar \lambda} 
\tripla R \tripla_{a'} \tripla h \tripla_{a'}  \e^{-\bar \lambda t}$$
This inequality implies the existence of
$$\lim_{t\to+\infty} h(t) = h_{\infty},$$
with $h_{\infty}\in \Cal X_{{\bar \lambda},\gamma}$ because 
$h\in \Cal B_{a',\gamma}$. Being $\gamma\ge 3$, the norm $\|h\|_{0,\gamma}$ dominates
the sup norm in $\vartheta$ and $\omega$, then 
 $h(t)$ converges exponentially fast to $h_{\infty}$
in the sup norm.\\ \\
\noindent
{\it Remark.} In the analysis carried out in this work the term $\hat{h}_0(t,\eta)=\frac{\hat{g}(\eta)}{\sqrt{2\pi}}$ can be separated from the other Fourier modes: in the Prop. \ref{propo:stimar} and its following, we can bound separately the zero and nonzero modes, so that, being more careful in the estimates, it is true that
$$\tripla R \tripla_a \le C \left|\!\left|f_0-\frac{g}{2\pi}\right|\!\right|_{\lambda_0,\gamma} + 
C\mu \tripla g\tripla_{a}\tripla R \tripla_a  +C \mu \tripla R \tripla_a\,
\left|\!\left|\!\left| h-\frac{g}{2\pi}\right|\!\right|\!\right|_a.$$
$$\left|\!\left|\!\left| h-\frac{g}{2\pi}\right|\!\right|\!\right|_a
\le C\left|\!\left|f_0-\frac{g}{2\pi}\right|\!\right|_{\lambda_0,\gamma}+ 
C \mu \tripla R\tripla_a\,\left|\!\left|\!\left| h-\frac{g}{2\pi}\right|\!\right|\!\right|_a.$$

Using these estimates, to prove our main theorem we need that $\mu$
is small only w.r.t. $\tripla g\tripla_a$, so 
we can keep the interaction fixed and take
small only the deviation from the constant phase density. In this way
we obtain, with a simple proof, a result similar to \cite{giacomin},
in which the authors prove that, for a sufficiently small perturbation
of the constant phase density, the solution of the equation
\eqref{noneq} is asymptotically a free flow and its $R$ vanishes.  The
threshold for $\mu$ that we obtain here could be non optimal, in fact
we do not take advantage from a detailed linear analysis as in
\cite{ldsobolev} and \cite{giacomin}.

\endproof

\bibliographystyle{plain}

\begin{thebibliography}{10}

\bibitem{bedrossian}
Bedrossian J. and Masmoudi N..
\newblock Inviscid damping and the asymptotic stability of planar shear flows
  in the 2d euler equations, 2013.


\bibitem{kuramoto2014}
Benedetto D. and Caglioti E. and Montemagno U.
\newblock On the complete phase synchronization for the Kuramoto model in the
  mean-field limit.
\newblock {\em arXiv:1407.6551}, 2014.
\newblock {(To appear in {\em Comm. Math. Sci.})} 

\bibitem{dephback}
Benedetto D. and Caglioti E. and Montemagno U.
\newblock Dephasing of the solutions of the kinetic Kuramoto model 
towards a fixed asymptotically free state.
\newblock  {\em arXiv:1411.6304}, 2014.
\newblock {(To appear in {\em Rend. Mat. and Appl.})}

\bibitem{caflish}
R. E. Caflish 
\newblock A simplified version of the abstract Cauchy-Kowalewski Theorem
with weak singularities
\newblock {\em Bull. Am. Math. Soc.} 23(2):495--500, 1990.

\bibitem{cm}
Caglioti E. and Maffei C.
\newblock Time asymptotics for solutions of {V}lasov-{P}oisson equation in a
  circle.
\newblock {\em J. Statist. Phys.}, 92(1-2):301--323, 1998.

\bibitem{carrillo}
Carrillo J.A., Choi Y.P., Ha S.Y., Kang M.J., and Kim Y.
\newblock Contractivity of transport distances for the kinetic kuramoto
  equation.
\newblock {\em Journal of Statistical Physics}, 156(2):395--415, 2014.

\bibitem{MR2897541}
Choi Y.P., Ha S.Y., Jung S., and Kim Y.
\newblock Asymptotic formation and orbital stability of phase-locked states for
  the {K}uramoto model.
\newblock {\em Phys. D}, 241(7):735--754, 2012.

\bibitem{dong}
Dong J.G. and Xue X..
\newblock Synchronization analysis of {K}uramoto oscillators.
\newblock {\em Commun. Math. Sci.}, 11(2):465--480, 2013.

\bibitem{ldsobolev}
Faou E. and Rousset F.
\newblock Landau damping in Sobolev spaces for the Vlasov-HMF model.
\newblock {\em arXiv:1403.1668}, 2014.

\bibitem{giacomin}
Fernandez B., Varet D.G., and Giacomin G.
\newblock Landau damping in the kuramoto model, 2014.

\bibitem{HHK}
Ha S.Y., Ha T., and Kim J.H.
\newblock On the complete synchronization of the {K}uramoto phase model.
\newblock {\em Phys. D}, 239(17):1692--1700, 2010.

\bibitem{ldexp}
Bedrossian J., Masmoudi N., and Mouhot C.
\newblock Landau damping: paraproducts and gevrey regularity.
\newblock {\em arXiv:1311.2870}, 2013.

\bibitem{bonilla}
Acebr\'on J.A., Bonilla L.L., P\'erez~Vicente C.J., Ritort F., and Renato
  Spigler.
\newblock The Kuramoto model: A simple paradigm for synchronization phenomena.
\newblock {\em Rev. Mod. Phys.}, 77(137), 2005.

\bibitem{kuramoto}
Kuramoto Y..
\newblock Self-entrainment of a population of coupled non-linear oscillators.
\newblock In Huzihiro Araki, editor, {\em International Symposium on
  Mathematical Problems in Theoretical Physics}, volume~39 of {\em Lecture
  Notes in Physics}, pages 420--422. Springer Berlin Heidelberg, 1975.

\bibitem{lancellotti}
Lancellotti C..
\newblock On the {V}lasov limit for systems of nonlinearly coupled oscillators
  without noise.
\newblock {\em Transport Theory Statist. Phys.}, 34(7):523--535, 2005.


\bibitem{mirolloor}
Mirollo R.E.
\newblock The asymptotic behavior of the order parameter for the infinite-N Kuramoto model. \newblock {\em Transport Theory Statist. Phys.}, 34(7):523--535, 2005.


\bibitem{villani}
Mouhot C. and Villani C.
\newblock On landau damping.
\newblock {\em Chaos: An Interdisciplinary Journal of Nonlinear Science}, 2012.

\bibitem{strogatz2000}
Strogatz S.H.
\newblock From {K}uramoto to {C}rawford: exploring the onset of synchronization
  in populations of coupled oscillators.
\newblock {\em Phys. D}, 143(1-4):1--20, 2000.
\newblock Bifurcations, patterns and symmetry.

\bibitem{ldstrogatz}
Strogatz S.H., Mirollo R.E., and Matthews P.C.
\newblock Coupled nonlinear oscillators below the synchronization threshold:
  Relaxation by generalized landau damping.
\newblock {\em Phys. Rev. Lett.}, 68:2730--2733, May 1992.

\end{thebibliography}

\end{document}